# Dry Dilution Refrigerator for Experiments on Quantum Effects in the Microwave Regime

A. Marx, J. Hoess, and K. Uhlig

Walther-Meißner-Institut
Garching, Germany 85748

## ABSTRACT

At the Walther-Meißner-Institut (WMI), a new cryogen-free $^3$He/$^4$He dilution refrigerator (DR) has been completed; the cryostat will be employed to cool experiments on superconducting quantum circuits for quantum information technology and quantum simulations. All major components have been made at the WMI. The DR offers lots of space at the various stages of the apparatus for microwave components and cables. E. g., the usable space at the mixing chamber has a height of more than 60 cm and a diameter of 30 cm (mixing chamber mounting plate). To cool cables and cold amplifiers, the DR is equipped with a separate $^4$He-1K-loop which offers a cooling power of up to 100 mW near 1K. The refrigeration power of the still is 18 mW at 0.9 K; the diameter of its mounting plate is 35 cm.

The cryostat rests in an aluminum trestle on air springs to attenuate building vibrations. It is precooled by a Cryomech PT410-RM pulse tube cryocooler (PTC) which is mechanically decoupled from the vacuum can of the cryostat by a bellows assembly. The two stages of the PTC are thermally connected to the DR via copper ropes. There are no nitrogen cooled traps with this DR to purify the gas streams of the $^3$He and $^4$He loops; instead, charcoal traps are mounted inside the DR at the first stage of the PTC. The dilution unit has three heat exchangers; its base temperature is 11 mK and its cooling power is 300 µW at 100 mK.

## INTRODUCTION

In low temperature research, when temperatures below 0.3 K are needed and experimental times exceed about 24 hours, $^3$He/$^4$He dilution refrigerators (DR) are indispensable and always utilized for cooling; experimental times with these refrigerators are sometimes as long as a year. In the years past, cryogen-free ("dry") DRs have more and more replaced traditional cryostats with liquid helium precooling because of their ease of operation and cost-effectiveness[1]. Besides, the amount of space for experiments in the mixing chamber region is usually quite big with dry DRs. Diameters of the mounting plate of the mixing chamber are usually around 30 cm with standard commercial DRs, but much larger diameters have also been reported[2]. Most dry DRs have only one big vacuum can. Therefore, to precool the dilution refrigeration unit from 300 K to ~10 K, exchange gas precooling is not possible. Instead, precooling of the various stages of the DR is accomplished by a separate precool loop where helium gas is circulated, or by gas switches. In our DR, a precool loop is installed.

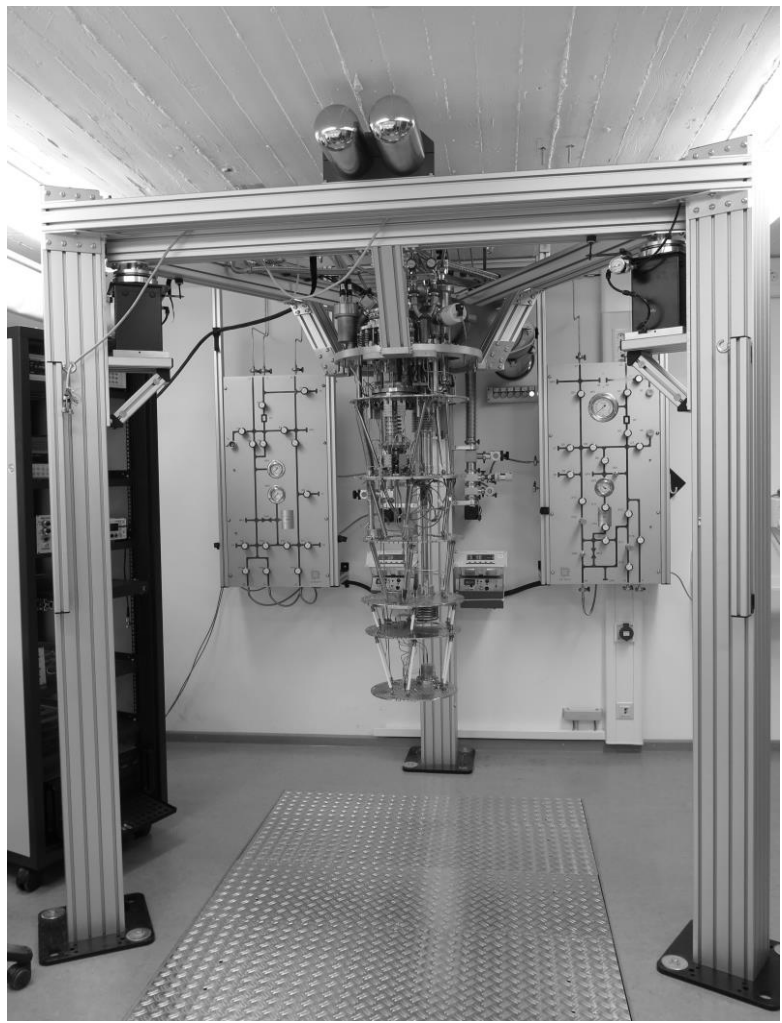

**Figure 1.** Photo of the open cryostat. It is mounted in the center of a tripod trestle. On either side of the DR, the gas handling boards are mounted to the wall in the back. All shields have been lowered into a pit underneath the cryostat.

In dry DRs, in order to heat-sink cables and to cool amplifiers at a temperature near 1K, the cooling power of the still $Q_{st}$ has to be used. $Q_{st}$ is proportional to the $^3$He flow of the DR and therefore quite limited. In addition, $Q_{st}$ is innately smaller with dry DRs compared with DRs with liquid helium precooling as there is no $^4$He condensation stage ("pot") with dry DRs; here, the returning $^3$He gas of the dilution circuit has to be cooled and liquefied in a heat exchanger (hx) which is cooled by the exhaust $^3$He gas of the still and by the refrigeration power of the still.

In our newly built dry DR where experiments on superconducting quantum circuits are to be conducted, a separate $^4$He-1K-loop was installed. It provides refrigeration powers of up to 100 mW near 1 K, several times higher than the refrigeration power of the still. It is also used to condense the $^3$He flow of the DR, and therefore the cooling power of the still is higher than that of a standard dry DR without 1K-loop. Extensive preparative work had been done on this subject in a test cryostat, before.[3]

**SETUP OF THE APPARATUS**

The cryostat rests in a tripod trestle which is equipped with air springs[4] to keep mechanical vibrations of the building away from the DR. Two separate gas handling boards for the DR and for the 1K-circuit are mounted to a wall next to the cryostat (Fig. 1). These boards are of the manually operated type, except for two automatic pressure relief valves which dump the helium gas into storage tanks in case of a power outage or of a block in a condensation line. The pumps to circulate the $^3$He and $^4$He of the two cryo-loops and a compressor to operate the precool loop are located in a separate machine room next to the refrigerator. The helium storage tanks are also placed in this room. For the DR, a turbo pump (Pfeiffer, HiPace 700) and an oil-free scroll pump (Edwards, XDS 35i) are available to circulate the $^3$He/$^4$He mash. The pumping line between the turbo pump and the cryostat is 3 m long and consists mostly of a bellows tube (150 mm i.d., 2.5 m length) to block the vibrations of the turbo pump. For operation of the 1K-loop we used a rotary pump (Edwards E2M80), so far. An alternative we plan to test next are two rotary pumps (Pfeiffer DUO20M) in parallel which are also available; they come with magnetic couplings between motor and pump, so air leaks of the shaft sealing gasket are avoided. A small membrane compressor (Hyco)[5] is necessary to circulate $^4$He in our precool loop during cooldowns. For its operation, the $^4$He gas and the gas handling board of the 1K-circuit can be used.

The vacuum can of the cryostat and both heat shields of the PTC are from aluminum (weight: 17.7 kg and 9.9 kg), whereas the still heat shield is from copper (weight: 21.4 kg). To open the cryostat, the vacuum can and the heat shields can be lowered into a pit (depth: 2.3 m).

The returning $^3$He/$^4$He gas flow of the dilution circuit and the $^4$He stream of the 1K-stage are first cooled and purified in two charcoal filters which are linked to the 1$^{st}$ stage of the PTC via copper braided ropes. Then both gas streams are cooled in identical hxs at the first stage, at the regenerator of the second stage and at the second stage of the PTC. The hxs at the first and second stage are made from little copper cylinders where CuNi capillaries have been wound around and soldered to; the cylinders are bolted to the flanges of the PTC. The hx mounted to the second regenerator consists of 10 little pipe sections where the two inlet lines of the gas circuits are soldered to. The pipe sections are clamped to the regenerator tube.[3] When the helium streams leave the second hx attached to the PTC, they have a typical temperature of ~ 2.7 K. Here, the $^4$He flow of the 1K-loop is liquid in normal operation, whereas the $^3$He flow of the DR, depending on the pressure, can be liquid or gaseous.

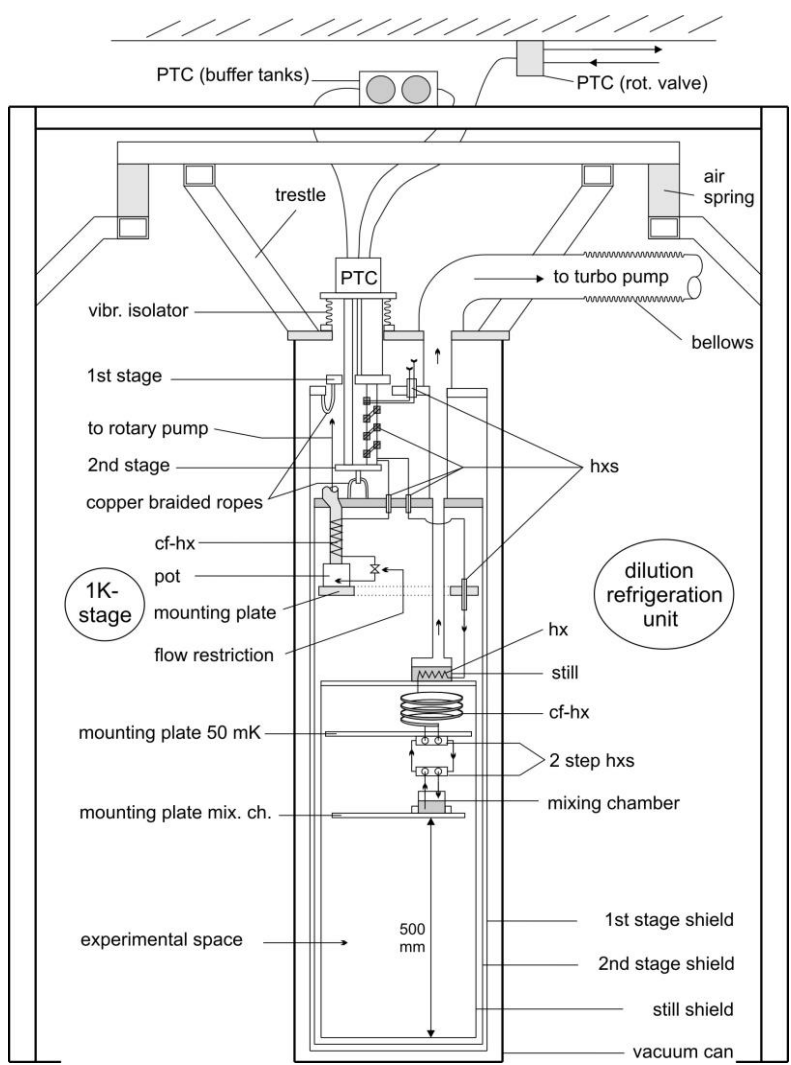

**Figure 2.** Cross section of the DR with 1K-loop. The 1K-stage is on the left side of the figure; the dilution unit (right side) consists of the standard components of cryogen-free DRs (for details see text). For clarity, the structure support posts in the dilution unit, the charcoal traps and the precool loop are not shown in the figure.

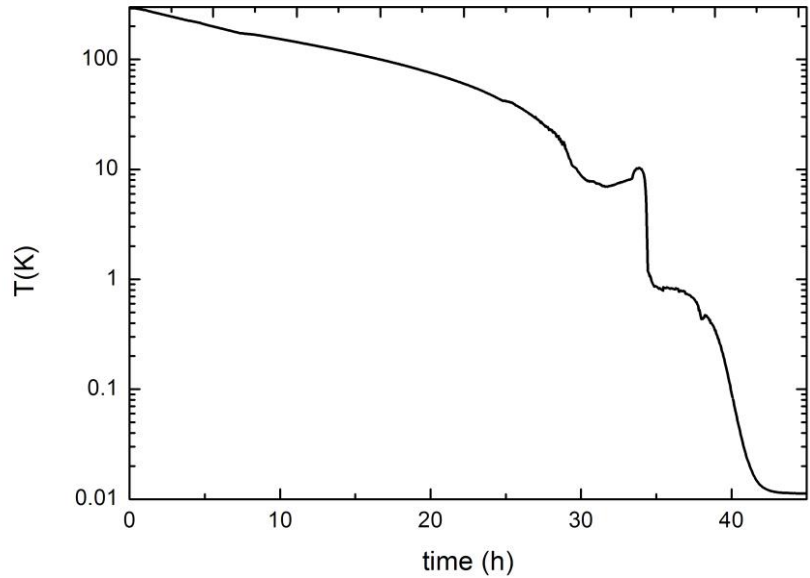

**Figure 3.** Cooldown curve of the DR. Helium circulation in the precool loop was stopped after 30 hours and the precool loop was pumped for 2 hours which results in a slight temperature increase. The adjacent temperature drop from 10 K to 1 K is caused by the starting of the 1K-loop. Finally, the DR cools the mixing chamber mounting plate to ~ 10 mK.

Our precool circuit is very similar to the one described in detail in other work by Batey et al.[7]. A helium gas stream is cooled in hxs by the two stages of the PTC to the temperature of the second stage; this cold helium stream is thermally connected by hxs to the mounting plates of the 1K-stage, the still, the 50 mK plate of the DR and the mixing chamber. The enthalpy of the outflowing helium gas is utilized in a counterflow hx to precool the back-streaming helium gas. There is no filter in this gas loop. The inlet pressure of the helium stream is 0.4 MPa at the beginning of a cooldown; this pressure decreases gradually with decreasing temperatures of the cryostat. A typical cooldown from room temperature to the base temperature of the DR (~ 10 mK) is given in Fig. 3. With the precool loop alone, a final precool temperature below 10 K is reached. After the $^4$He flow is stopped, the precool loop has to be evacuated thoroughly so a superfluid film which could thermally short the DR, cannot develop on further cooling.

## $^4$HE-1K-STAGE

Below the second stage of the PTC, the 1K-loop consists of a counterflow hx[8], a flow restriction and a vessel where the liquid helium can accumulate (Fig. 2); the volume of the vessel is 54 cm$^3$. Thus the 1K-stage is identical to a JT-stage, with the exception that liquid helium is fed into its inlet. The bottom of the vessel is bolted to a copper mounting plate (D = 36 cm) where many electric lines and coax cables can be fed through and heat-sunk. A hx where the $^3$He of the dilution circuit is condensed is also affixed to this plate. The flow restriction is made from a piece of capillary (0.1 mm i.d., l = 15 cm); so the flow impedance is fixed. A heater and a RuOx thermometer[9] are attached to the mounting plate so the refrigeration power of the 1K-stage could be measured (Fig. 4).

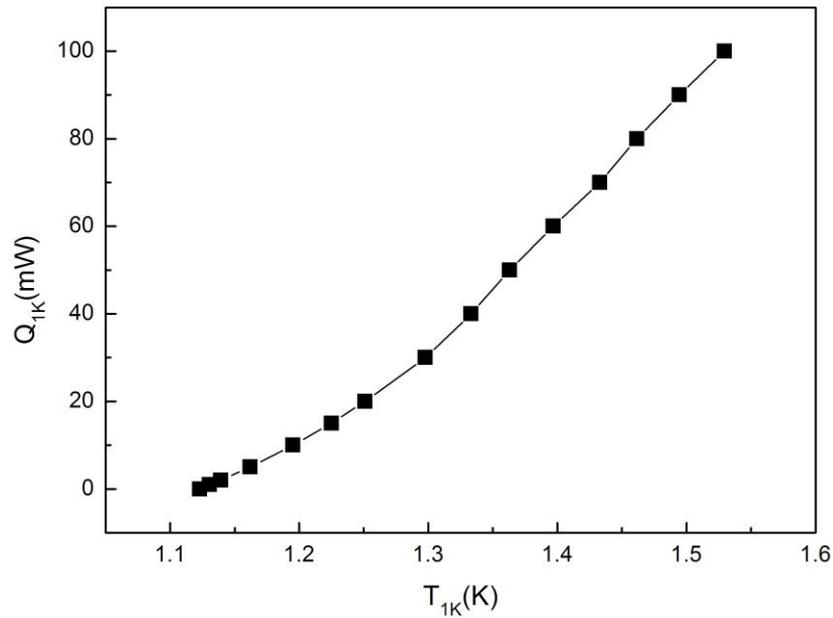

**Figure 4.** Refrigeration power $Q_{1K}$ of the 1K-stage as a function of its temperature. The base temperature is just above 1.1 K when it is operated with a rotary pump. At the highest $Q_{1K}$ of 100 mW the temperature of the mounting plate is 1.55 K.

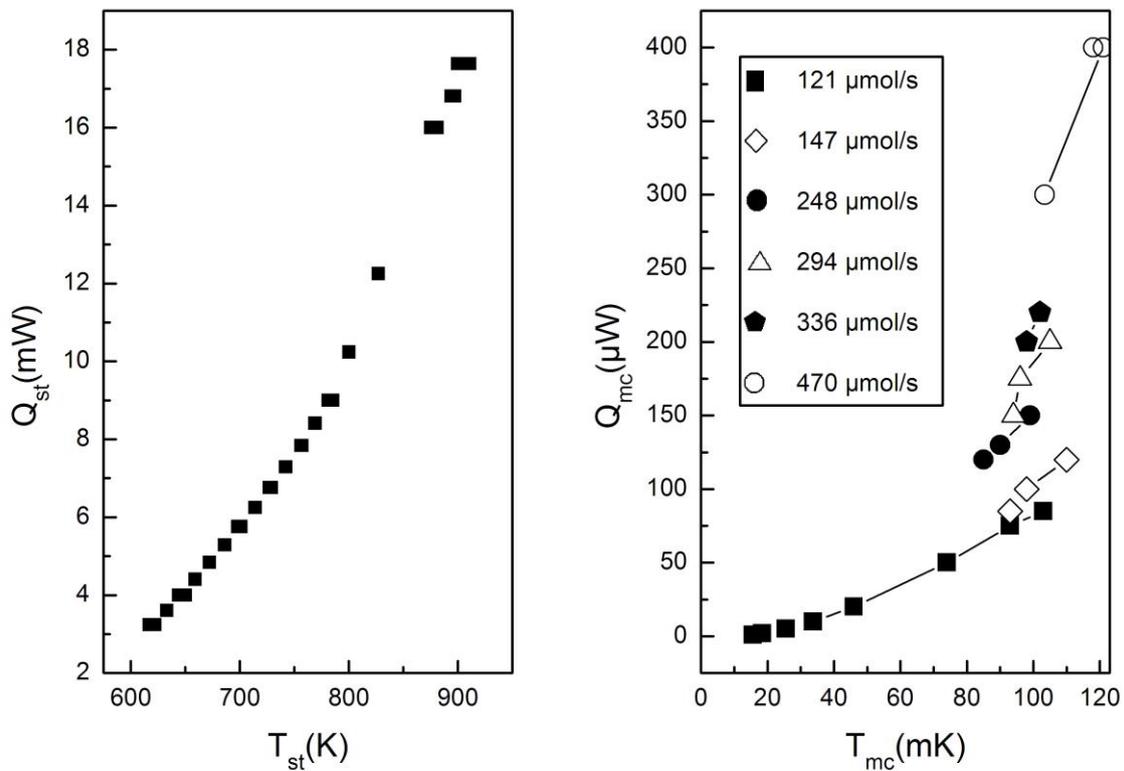

**Figure 5.** Cooling power of the still ($Q_{st}$) and of the mixing chamber ($Q_{mc}$) as functions of their respective temperatures; $Q_{st}$ can reach 18 mW at 900 mK. $Q_{mc}$ is given for six different $^3$He flow rates. The lowest temperature of the mixing chamber of 11 mK is reached with a small $^3$He circulation rate. At a mixing chamber temperature of 100 mK, the cooling power of the mixing chamber is 0.3 mW at the highest $^3$He circulation rate.

## DILUTION REFRIGERATION CIRCUIT

The dilution refrigeration unit was made at the WMI, but is fairly standard. It consists of a still chamber (V = 53 cm$^3$), a continuous hx, 2 step hxs, and a mixing chamber with a large silver sponge to couple thermally the helium liquid to a mounting plate (Fig. 2). The still chamber is bolted to a mounting plate of 35.2 cm diameter where electrical leads and coax lines can be attached and heat sunk. The continuous hx is similar to the one described in Ref. 10. The two step exchangers are made from silver blocks with silver sponges inside for thermal contact between the liquid helium and the hx body. Quite similar ones have been described in Ref. 11. The hotter one of these step hxs is bolted to a mounting plate of 33 cm diameter. With the mixing chamber close to its base temperature, the temperature of this mounting plate is ~ 50 mK. The mixing chamber has an inner volume of 63 cm$^3$. There are 76 silver wires welded into its bottom plate where each wire carries a silver sponge of 1.2 cm$^3$ volume. The bottom plate of the mixing chamber is bolted to a mounting plate of 30 cm diameter. The mounting plates of the various stages of the DR are held by a structure made of G10 (glass fiber reinforced epoxy) pipes. The refrigeration powers of the still and of the mixing chamber are given in Fig. 5. Temperatures were measured with RuOx resistance thermometers[9], and $^3$He flow rates were determined from the inlet pressure of the fore pump and its corresponding pumping speed.

## SUMMARY

In our report we describe a newly constructed cryogen-free DR; it is equipped with an additional 1K-loop which provides a refrigeration power of up to 100 mW. The condensation rate of the $^3$He/$^4$He mixture prior to an experiment has been improved by a factor of two with the use of the 1K-stage. The experimental setup, the cooldown of the DR and the refrigeration powers of the 1K-loop, of the still and of the mixing chamber of the DR are presented in the paper.

## ACKNOWLEDGEMENT

This work is supported by the German Research Foundation through SFB 631, the German Excellence Initiative through NIM and EU projects CCQED and PROMISCE.